\documentclass[aps,prl,twocolumn,groupedaddress,showpacs]{revtex4-1}
\usepackage{dcolumn} 
\usepackage{bm} 
\usepackage{amsthm}
\usepackage{latexsym}
\usepackage{float}
\usepackage{amsfonts}
\usepackage{amsmath}
\usepackage{amssymb}
\usepackage{color,graphicx}
\usepackage{booktabs}
\usepackage{epstopdf}
\newcommand{\dprime}{\prime\prime}
\begin{document}
\title{Ultrafast Control of Crystal Structure in a Topological Charge-Density-Wave Material}
\author{Takeshi Suzuki$^{1,*}$}
\author{Yuya Kubota$^{2}$}
\author{Natsuki Mitsuishi$^{3}$}
\author{Shunsuke Akatsuka$^{4}$}
\author{Jumpei Koga$^{4}$}
\author{Masato Sakano$^{4}$}
\author{Satoru Masubuchi$^{5}$}
\author{Yoshikazu Tanaka$^{2}$}
\author{Tadashi Togashi$^{2,6}$}
\author{Hiroyuki Ohsumi$^{2}$}
\author{Kenji Tamasaku$^{2}$}
\author{Makina Yabashi$^{2,6}$}
\author{Hidefumi Takahashi$^{7,8}$}
\author{Shintaro Ishiwata$^{7,8}$}
\author{Tomoki Machida$^{5}$}
\author{Iwao Matsuda$^{1}$}
\author{Kyoko Ishizaka$^{3,4}$}
\author{Kozo Okazaki$^{1,*}$}
\affiliation{$^1$Institute for Solid State Physics, The University of Tokyo, Kashiwa, Chiba 277-8581, Japan\\
$^2$ RIKEN SPring-8 Center, 1-1-1 Kouto, Sayo, Hyogo 679-5148, Japan\\
$^3$ RIKEN Center for Emergent Matter Science (CEMS), Wako 351-0198, Japan\\
$^4$ Quantum-Phase Electronics Center and Department of Applied Physics, The University of Tokyo, Tokyo 113-8656, Japan\\
$^5$ Institute of Industrial Science, The University of Tokyo, Meguro-ku, Tokyo 153–8505, Japan\\
$^6$ Japan Synchrotron Radiation Research Institute (JASRI), 1-1-1 Kouto, Sayo, Hyogo 679-5198, Japan\\
$^7$ Division of Materials Physics and Center for Spintronics Research Network (CSRN), Graduate School of Engineering Science, Osaka University, Toyonaka, Osaka 560-8531, Japan\\
$^8$ Spintronics Research Network Division, Institute for Open and Transdisciplinary Research Initiatives, Osaka University, Yamadaoka 2-1, Suita, Osaka 565-0871, Japan
}
\date{\today}

\begin{abstract}
Optical control of crystal structures is a promising route to change physical properties including topological nature of a targeting material.
Time-resolved X-ray diffraction measurements using the X-ray free-electron laser are performed to study the ultrafast lattice dynamics of VTe$_2$, which shows a unique charge-density-wave (CDW) ordering coupled to the topological surface states as a first-order phase transition.
A significant oscillation of the CDW amplitude mode is observed at a superlattice reflection as well as Bragg reflections.
The frequency of the oscillation is independent of the fluence of the pumping laser, which is prominent to the CDW ordering of the first-order phase transition.
Furthermore, the timescale of the photoinduced 1$T^{\dprime}$ to 1$T$ phase transition is independent of the period of the CDW amplitude mode.
\end{abstract}

\maketitle

Engineering crystal structures is one of the direct ways to change electronic, optical, and mechanical properties in solid state materials.
Many techniques of a structural control have been developed including strain \cite{Kim_2018}, nanostructuring \cite{Marconnet_2013}, heterostructure layer stacking \cite{Liao_2019}, and twistronics \cite{Cao_2018a, Cao_2018b}.
While these approaches are very powerful, investigations are limited to equilibrium states, where all the degrees of freedom are thermally balanced.
On the other hand, the combination of photoexcitation by ultrafast pulse laser and various probing techniques has enabled us to study nonequilibrium states, where a specific subsystem is selectively excited, by which a lot of exotic phenomena have been reported \cite{Basov_2017}.
Ultrafast optical pulses have also served as powerful tools to engineer crystal structures, by which a variety of phases, including superconductivity \cite{Fausti_2011, Mankowsky_2014}, ferroelectricity \cite{Nova_2019, Li_2019}, and magnetism \cite{Radu_2011}, have been found to emerge.

Recently, controlling topological properties of a material by light has attracted enormous interest because a topological insulator has robust metallic surface states against impurities or disorder \cite{Roushan_2009, Hasan_2010}, which is preferable for optical switching.
While many approaches to the ultrafast topological control have been demonstrated by such as Floquet engineering \cite{Oka_2009, Wang_2013, Mahmood_2016, Mclver_2020}, controlling crystal structures with light has also been employed to manipulate topological properties \cite{Sie_2019, Disa_2021}.
This methodology is also applicable to charge-density-wave (CDW) materials having topological properties.
A great advantage of using CDW materials is the feasibility of flexibly tuning their properties by external stimuli such as physical \cite{Sipos_2008, Kusmartseva_2009} or chemical pressure \cite{Ru_2008, Brouet_2008}, and electric \cite{Lee_1979, DiCarlo_1993, Adelman_1995} or magnetic fields \cite{Chang_2016}.

For the optical control of CDW materials, a lot of studies have been reported mainly in the context of collective-mode excitations or photoinduced phase transitions.
Since a CDW phase is a coupled phase between charge and lattice, multiple ultrafast probing techniques have been employed to track different degrees of freedom, including optical pump-probe spectroscopy \cite{Demsar_1999, Demsar_2002, Yusupov_2010}, time- and angle-resolved photoemission spectroscopy \cite{Perfetti_2006, Schmitt_2008, Hellmann_2012, Makler_2021}, ultrafast electron diffraction (UED) \cite{Eichberger_2010, Erasmus_2012, Haupt_2016, Kogar_2020}, and time-resolved X-ray diffraction (TRXRD) \cite{Mohr-Vorobeva_2011, Huber_2014, Laulhe_2017, Trigo_2019, Burian_2021}.
These reports have highlighted second-order CDW phase transition systems except for 1$T$-TaS$_2$.
On the other hand, a specific phase is preserved against the external turbulence within the threshold in a first-order phase transition.
Such a \textit{robustness} of the phase is great advantage over a second-order phase transition, especially for the application of ultrafast devices such as an optical switch or memory medium.
In terms of fundamental physics, kinetics of first-order photoinduced phase transitions shows many attractive phenomena such as nucleation \cite{Sternbach_2021} or percolative dynamics \cite{Teitelbaum_2019}.
Thus, it is highly desired to study in detail the dynamical features of a first-order photoinduced CDW phase transition system having intriguing profiles.

In this Letter, we report ultrafast lattice dynamics in VTe$_2$, which has a first-order CDW phase transition and the topological surface states coupled to the CDW phase, by performing the measurements of TRXRD using the X-ray free-electron laser (XFEL).
We directly observe the CDW amplitude mode in VTe$_2$ as a significant temporal oscillation, 1.5 THz in frequency, of a superlattice reflection.
We further study CDW melting dynamics by the pump-fluence dependent measurements.
Through the fitting analysis, we successfully reveal the distinctive features of a first-order CDW phase transition.
They are quite different from many other CDW materials of second-order phase transitions.
Our findings provide detailed guidelines for the crystal structure control of first-order CDW materials and an important step towards the manipulation of topological surface states.

\begin{figure}[!t]
\includegraphics{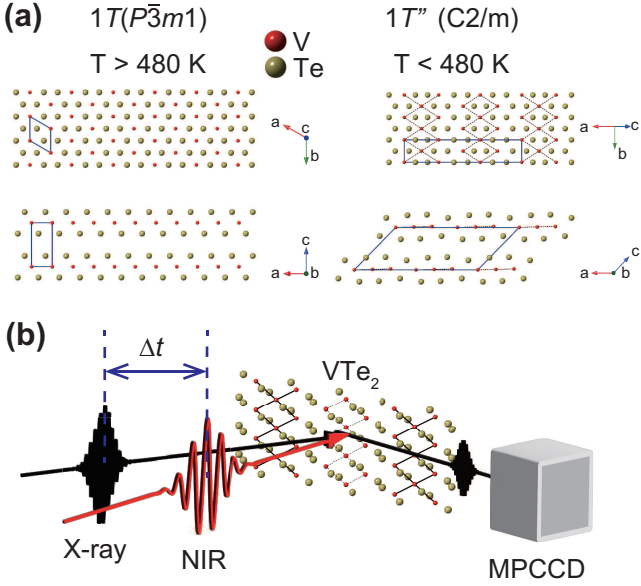}
\caption{
(a) Crystal structure of VTe$_2$.
Above the phase transition temperature of 480 K, VTe$_2$ displays the 1$T$ ($P3\overline{m}1$) structure while below 480 K it shows the 1$T^{\dprime}$ ($C2/m$) structure.
The double-zigzag bonds of the V atoms in the 1$T^{\dprime}$ phase are shown as black-dashed lines and a unit cell for each phase is shown as blue-solid box. 
(b) Schematic illustration of the time-resolved X-ray diffraction measurement.
The delay time between the near-infrared (NIR) pump and the hard X-ray probe is shown as $\Delta t$.
The diffracted signal is detected by the multi-port charge-coupled-devises (MPCCD).
}
\label{fig1}
\end{figure}

Figure 1(a) shows the crystal structure of VTe$_{2}$ in the 1$T$ ($P\overline{3}m1$, \#164 trigonal) metallic and 1$T^{\dprime}$ ($C2/m$, \#12 monoclinic) CDW phases, where a unit cell for each phase is shown as blue-solid box.
With lowering temperature, VTe$_{2}$ displays the CDW phase transition at 480 K, accompanied by a formation of $3 \times 1 \times 3$ superstructure characterized by the double-zigzag chain structure for the V atoms along the $b$ direction, shown as the black-dashed lines in Fig. 1(a) \cite{Bronsema_1984}.
One of the most striking signatures is the significantly large contraction of the V-V bonds ($\sim$ 9.1 \%), which is larger than Ta-Ta bonds in the well-known CDW material, 1$T$-TaS$_2$ ($\sim$ 7.0 \% \cite{Brouwer_1980}).
The hysteresis of resistivity in VTe$_2$ \cite{Ohtani_1981} also confirms that this CDW transition in VTe$_{2}$ is of first-order.
Furthermore, the recent study of angle-resolved photoemission spectroscopy revealed an intimate relationship between the topological surface states and CDW order \cite{Mitsuishi_2020}, where two of three Dirac surface states at the boundary of the 1st Brilluin zone in the 1$T$ metallic phase were found to disappear in the 1$T^{\dprime}$ CDW phase.
For the CDW melting dynamics, the UED measurements were performed to directly observe the lattice dynamics of VTe$_2$, and successfully observed the acoustic phonons \cite{Nakamura_2018, Nakamura_2020}.
However, due to the limitation of the time resolution of $\sim$ 2 ps, the initial dynamics of the CDW melting has remained uncovered.
The optical pump-probe measurements with the sufficient time resolution were performed to observe the multiple coherent phonons \cite{Tanimura_2022} while a direct insight into the lattice dynamics is still missing.

To directly track the lattice dynamics with high temporal resolution, we perform the measurements of TRXRD at BL3 of the XFEL facility, SPring-8 Angstrom Compact free-electron LAser (SACLA) \cite{Ishikawa_2012} as schematically shown in Fig. 1 (b).
The experimental setup and bulk sample preparation are written in Supplementary Materials for details \cite{Suppl}.
To minimize the penetration-length mismatch between the NIR pump and X-ray probe (23 nm \cite{Tanimura_2022} and $\sim$10 $\mu$m, respectively, for VTe$_2$), we prepared the samples with the thickness of $\sim$40 nm using mechanical exfoliation by transparent adhesive tape and dry transfer onto the Si$_3$N$_4$ membrane \cite{Suppl}.
The thickness of the sample was measured by both atomic force microscopy and acoustic wave measurements \cite{Suppl}.
Integrated intensity of the diffracted image $I_{hkl}$ is given by the structure factor $F_{hkl}$ as expressed by $I_{hkl} \propto \left| F_{hkl} \right|^2$, and the structure factor is determined by atomic positions in a unit cell as
\begin{equation}
F_{hkl} = \sum_{n} f_{n} \exp \left( i \bm{G}_{hkl} \cdot \bm{r}_{n} \right), 
\end{equation}
where $hkl$ is the Bragg reflection index, $f_n$ is the atomic form factor of the $n$th atom, $\bm{G}_{hkl}$ is the reciprocal vector, and $\bm{r}_{n}$ is the atomic position of the $n$th atom.
For the Bragg reflection indices in the main text, we use the practical monoclinic lattice \cite{Nakamura_2020, Suppl}.
The changes in diffraction \textit{angle} of Bragg reflections reflect the changes in lattice constants \cite{Suzuki_2021} while the changes in diffraction \textit{intensity} reflect the atomic displacements inside the unit cell.

\begin{figure}[!t]
\includegraphics{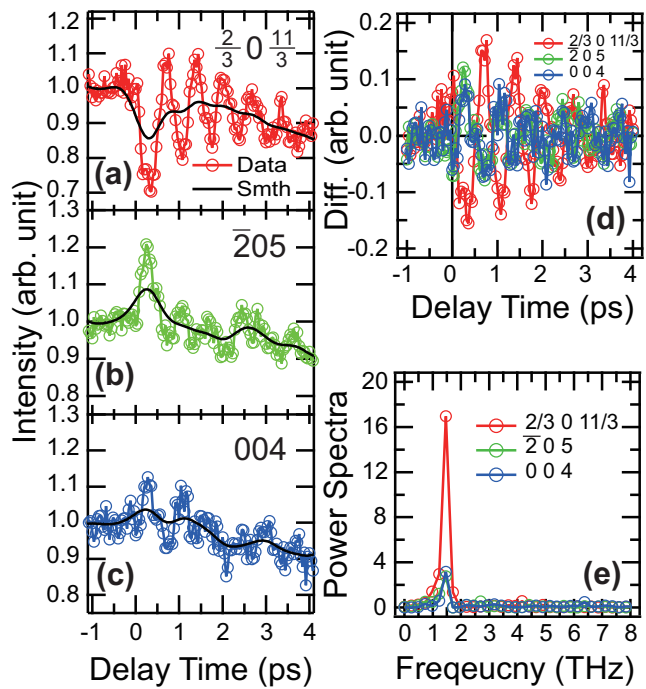}
\caption{
(a-c) Time-dependent $I_{\frac{2}{3}0\frac{11}{3}}$, $I_{\text{$\overline{2}$05}}$, and $I_{\text{004}}$, at the pump fluence of 0.8, 1.2, and 1.0 mJ/cm$^2$, respectively.
The data are shown as marks and the smoothing lines (smth) are shown as black solid lines.
(d) Oscillatory components of (a-c) deduced by subtracting smoothing lines from data.
(e) Power spectra of the oscillatory components (d).
}
\label{fig2}
\end{figure}

Figure 2(a) shows the time dependence of the superlattice reflection intensity $I_{\frac{2}{3}0\frac{11}{3}}$ obtained with the low pump fluence of 0.8 mJ/cm$^2$.
One can clearly see a significant oscillation.
$I_{\frac{2}{3}0\frac{11}{3}}$ is determined by the amount of the Peierls distortion, $x$, as $I_{\frac{2}{3}0\frac{11}{3}} \propto |x|^2$ because the structure factor of the superlattice reflection, $F_{\frac{2}{3}0\frac{11}{3}}$, is proportional to $x$ in the leading order \cite{Overhauser_1971}.
Therefore, the modulation of the superlattice reflection intensity can be attributed to the excitation of the CDW amplitude mode.
The previous studies using TRXRD on other CDW materials also linked the oscillation of superlattice reflections to the CDW amplitude modes \cite{Huber_2014, Trigo_2019, Burian_2021}.
To further confirm this assignment, we also measure the Bragg reflection intensities of $I_{\text{$\overline{2}$05}}$ and $I_{\text{004}}$ as shown in Figs. 2(b) and 2(c).
For both reflections, oscillations can be recognized.
To deduce the oscillatory components, we subtract the backgrounds given by the smoothed lines shown as the black-solid lines in Figs. 2(a)-2(c).
The subtracted data are shown in Fig. 2 (d).
Interestingly, the oscillation phases of the Bragg reflections are opposite to that of the superlattice reflection, \textit{i.e.} while the $I_{\frac{2}{3}0\frac{11}{3}}$ \textit{decreases} in the first quarter cycle, the $I_{\text{$\overline{2}$05}}$ and $I_{\text{004}}$ \textit{increase}.
Fourier transforms are performed for the subtracted data and the power spectra are shown in Fig. 2(e).
One can find a single-peak structure at 1.5 THz in all the diffraction intensities.
This frequency corresponds to $f_{\text{AM}}$ in Eq. (3) discussed later.

\begin{table}[h]
\caption{\label{tab:fonts}Structure factors of VTe$_2$ for 1$T$ and 1$T^{\dprime}$ phases}
\begin{ruledtabular}
\begin{tabular}{lccc}
& $F_{1T^{\dprime}}$ & $F_{1T}$ & $I_{1T}/I_{1T^{\dprime}}$ \\
$\overline{2}$05 & 280+3$i$ & 408+5$i$ & 2.12 \\
004 & 360+3$i$ & 514+6$i$ & 2.02
\end{tabular}
\end{ruledtabular}
\end{table}

To understand the opposite oscillation phase between the superlattice and Bragg reflections, we consider the origin of the intensity modulation of the Bragg reflections in terms of the structure factors.
While it is assumed that the photoinduced phase cannot be simply expressed using the structures of 1$T$ and 1$T^{\dprime}$ phases, here we approximate the structure factor in the photoinduced phase as the superposition of $F_{1T}$ and $F_{1T^{\dprime}}$.
$F_{\text{$\overline{2}$05}}$ and $F_{\text{004}}$ are calculated from Eq. (1) for both the 1$T$ and 1$T^{\dprime}$ phases \cite{Suppl}, and shown in Table I.
Both $|F_{\text{$\overline{2}$05}}|$ and $|F_{\text{004}}|$ for the 1$T$ phase are larger than those for the 1$T^{\dprime}$ phase.
This is consistent with the increasing behavior in the first quarter cycle and further supports the assignment of the CDW amplitude mode.
We note that the 1.5 THz coherent phonon was also confirmed by the previous optical pump-probe measurement \cite{Tanimura_2022}.

In order to investigate the CDW melting dynamics in VTe$_2$, we perform the measurements with higher pump fluence.
Figure 3 (a) shows the fluence-dependent dynamics of $I_{\frac{2}{3}0\frac{11}{3}}$.
While the CDW amplitude mode is clearly observed at low fluences, it becomes strikingly suppressed with increasing fluence.
Furthermore, the intensity drop immediately after the pump excitation increases, to the one tenth of the initial value at the highest fluence, which corresponds to the modulation from a structure of the 1$T^{\dprime}$ CDW phase to that close to the 1$T$ metallic phase.
Figure 3 (b) schematically shows these observations in terms of lattice dynamics in real space.
Under low fluences, the V atoms smoothly oscillate with small displacements to new stable positions, the same mechanism as the displacive excitation of coherent phonons (DECP) \cite{Zeiger_1992}.
Under high fluences, on the other hand, the V atoms are largely displaced from their original positions while the damping is strong enough to prevent multi-cycle oscillations.
At the large delay time of 3-4 ps, the intensity decreases with pump fluence and does not recover to the original value.
To show this behavior more clearly, we plot the corresponding values of the intensity as a function of pump fluence in Fig. 3(c).
Here $I_{\text{Slow}}$ represents the averaged intensity during $\Delta t$ = [3-4] ps.
$I_{\text{Slow}}$ decreases almost monotonically but slightly nonlinearly with pump fluence.
To highlight this behavior, we linearly fit the data at the low ($<$ 1.5 mJ/cm$^2$) and high fluences ($\ge$ 2.5 mJ/cm$^2$) and these lines are found to cross each other at 1.9 mJ/cm$^2$.
This value implies the threshold fluence of the CDW melting, and we intensively discuss the melting dynamics in the following paragraphs.

\begin{figure}[!t]
\includegraphics{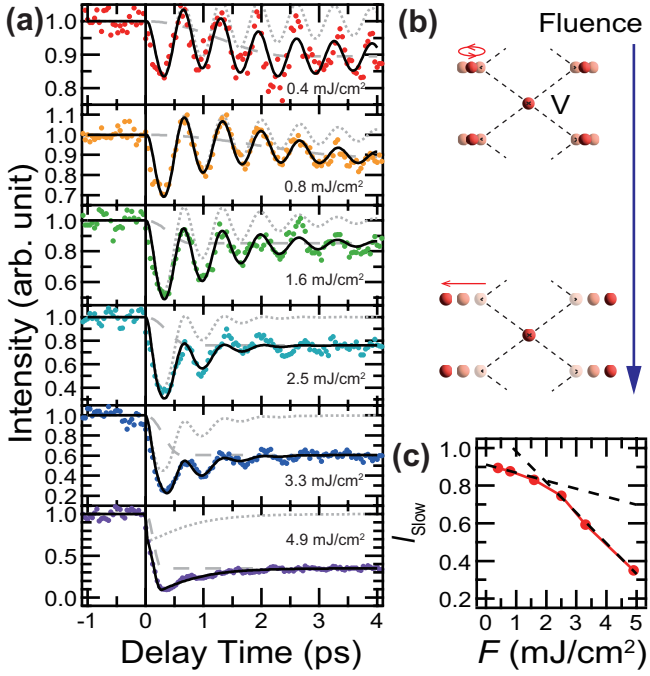}
\caption{
(a) Fluence-dependent dynamics of $ I_{\frac{2}{3}0\frac{11}{3}}$.
The fluences are indicated in each panel.
The data are shown as markers and the total fits are shown as black-solid lines.
The fits are composed of $I_1(t)$ and $I_2(t)$, and they are shown as gray-dashed and gray-dotted lines, respectively.
(b) Schematic illustration of the fluence-dependent CDW amplitude mode.
(c) Averaged intensity at $\Delta t$ = [3-4] ps, $I_{\text{Slow}}$, as a function of pump fluence $F$.
}
\label{fig3}
\end{figure}

\begin{figure}[!b]
\includegraphics{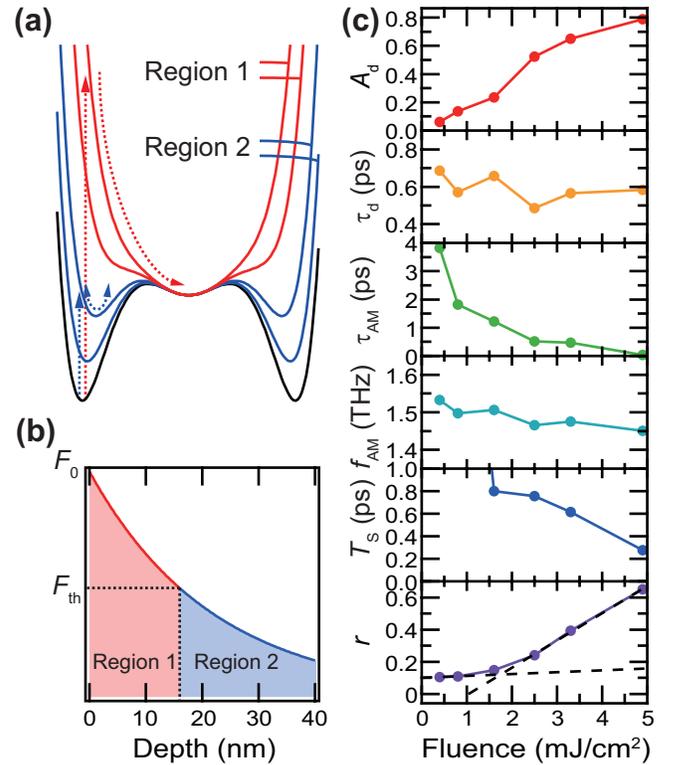}
\caption{
(a) Schematic picture of the energy potential and dynamics after photoexcitation in a first-order phase transition system.
(b) Schematic illustration of the two-component distribution of the excitation along the depth direction.
Regions 1 and 2 are the photoinduced metallic and CDW phases, respectively.
(c) Summary of the fitting parameters as a function of pump fluence.
}
\label{fig4}
\end{figure}

As we mentioned, VTe$_2$ is a first-order CDW phase transition material like 1$T$-TaS$_2$.
The photoinduced phase transition, then, can be pictured by the Landau potential for the first-order phase transition shown in Fig. 4(a) \cite{Binder_1987}.
Furthermore, the boundary between the 1$T^{\dprime}$ CDW and 1$T$ metallic phases is rather well separated by a so-called phase boundary, thus we assume the superlattice reflections during the photoinduced phase transition can be approximated as two spatially separate components.
Due to the attenuation of pump excitation, the photoinduced 1$T$ phase is triggered in a shallower region from the surface (Region 1) while the CDW phase remains in a deeper region from the surface (Region 2).
Figure 4(b) schematically shows this situation, where the incident pump fluence at the surface is denoted as $F_0$ and the threshold pump fluence for the CDW melting is denoted as $F_{\text{th}}$ estimated to be 1.9 mJ/cm$^2$ from Fig. 3(c).
This value nearly agrees with the previous study of 1.21 mJ/cm$^2$ \cite{Tanimura_2022}.
Strictly speaking, the photoinduced phase is assumed to be different from the 1$T$ or 1$T^{\dprime}$ phases that are defined in the equilibrium state.
Here, we approximately treat Regions 1 and 2 as 1$T$ or 1$T^{\dprime}$ phases, respectively.

We use the phenomenological two-component model composed of $I_{1} (t)$ and $I_{2} (t)$ for the intensity of superlattice reflections from Regions 1 and 2, respectively, shown below, in the same manner as the previous work on 1$T$-TaS$_2$ \cite{Laulhe_2017},
\begin{gather}
I_{1} (t) = \frac{1}{2} \left(1 + \cos \frac{\pi t}{T_S} \right) \qquad 0 < t < T_S \\
I_{2} (t) = \left( 1 + A_d \left[ \cos \left( 2 \pi f_{\text{AM}} t \right)e^{-t/\tau_{\text{AM}}} - e^{-t/\tau_{\text{d}}} \right] \right)^2.
\end{gather}
$I_{1} (t)$ is a sigmoid-shaped function, where $T_S$ is the CDW melting time.
$I_{2} (t)$ stands for the square of the atomic displacement for DECP dynamics.
$A_d$, $f_{\text{AM}}$, and $\tau_{\text{AM}}$ are the amplitude, frequency, and decay constant for the CDW amplitude modes, respectively, $\tau_{\text{d}}$ is the relaxation time back to the initial potential for Region 2.
We fit our experimental data by $I_{\text{Total}} (t) = r I_{1} (t) + (1-r) I_{2} (t)$, where $r$ is the volume fraction of Region 1.

The total fitting results are shown as black-solid lines in Fig. 3(a), where $I_{1} (t)$ and $I_{2} (t)$ are shown as gray-dashed and dotted lines, respectively.
Figure 4(c) shows the results of the fitting parameters.
As for the parameters for $I_2(t)$, $A_d$ increases, and $\tau_{\text{AM}}$ decreases with increasing fluence, corresponding to the large displacement and stronger damping schematically shown in Fig. 3(b).
The relaxation time $\tau_{d}$ does not change noticeably.
The small change of $f_{\text{AM}}$ is a quite remarkable behavior and a stark difference from the CDW materials of the second-order phase transitions \cite{Yusupov_2010, Huber_2014, Trigo_2019, Burian_2021, Makler_2021}, where significant changes of $f_{\text{AM}}$ as a function of pump fluence has been reported such as the dynamical slowing down \cite{Zong_2019} or the overshooting behavior \cite{Huber_2014}.
These observations have been explained in terms of dynamics on the ``Mexican-hat" potential in second-order phase transitions.
Thus our observation of $f_{\text{AM}}$ clearly marks the robust curvature of potential below the threshold in the first-order case (Fig. 4(a)).

As for the parameters for $I_1(t)$, we only show the result of $T_S$ between 0 and 1.0 ps, and the results at full scale are shown in Fig. S8 \cite{Suppl}.
$T_S$ slightly decreases with fluence up to 3.3 mJ/cm$^2$ and is larger than the half period ($\sim$ 0.3 ps) of the CDW amplitude mode of 1.5 THz.
This is a stark contrast to the previous work on 1$T$-TaS$_2$, where the CDW melting time was determined to the half period of the CDW amplitude mode \cite{Hellmann_2012}.
However, the more recent work on 1$T$-TaS$_2$ reported the independent value of the CDW melting time from the CDW amplitude mode \cite{Laulhe_2017}, which is the same conclusion as our current work.
With further increasing the fluence to 4.9 mJ/cm$^2$, $T_S$ significantly decreases.
This behavior is intuitively understood by considering the steeper gradient of the potential with the higher fluence shown in Fig. 4(a).
Although the microscopic mechanism determining the CDW melting time is still elusive, we can stress that it is not simply determined by the CDW amplitude mode. Further considerations of dynamical electron-electron and electron-phonon interactions should be crucial for a deeper understanding, because these intertwined interactions are expected to play an important role especially for materials such as 1$T$-TaS$_2$ and VTe$_2$.
Lastly, the volume fraction of Region 1, $r$, increases with pump fluence and shows a threshold-like behavior at 1.8 mJ/cm$^2$ from the linear fit at the low ($<$ 1 mJ/cm$^2$) and high fluences ($\ge$ 2.5 mJ/cm$^2$) as shown in Fig. 4(c), in close agreement with the value estimated in Fig. 3(c).

In summary, we investigated the CDW-amplitude-mode excitations and melting dynamics in VTe$_2$.
The characteristic behaviors of the first-order phase transition were revealed and our work provides valuable insight into the dynamical control of the first-order CDW materials.
Moreover, the topological surface states controlled via CDW phase in VTe$_2$ can further extend the potential of applications \cite{Mitsuishi_2020}, and lead to the ultrafast surface state switching.

\begin{acknowledgments}
This work was supported by Grants-in-Aid for Scientific Research (KAKENHI) (Grant No. JP18K13498, JP19H00659, JP19H01818, JP19H00651 JP18K14145, JP19H02623, JP20H01834, and JP21H05235) from the Japan Society for the Promotion of Science (JSPS), by JSPS KAKENHI on Innovative Areas “Quantum Liquid Crystals” (Grant No. JP19H05826), by the Center of Innovation Program from the Japan Science and Technology Agency, JST, the Research and Education Consortium for Innovation of Advanced Integrated Science by JST, and by MEXT Quantum Leap Flagship Program (MEXT Q-LEAP) (Grant No. JPMXS0118067246, JPMXS0118068681), Japan, by a CREST project (No. JPMJCR20B4).
This experiment was performed at BL3 of SACLA with the approval of the Japan Synchrotron Radiation Research Institute (JASRI) (Proposal No. 2022A8024).
The synchrotron radiation experiments were performed to evaluate the system at BL19LXU in SPring-8 with the approval of RIKEN (Proposal No. 20220049).
The authors would like to acknowledge the support of members of the SACLA and SPring-8 facilities.
The authors would like to acknowledge Dr. A. Nakamura for the fruitful discussion.
T. S. acknowledges the research grants from The Murata Science Foundation, The Hattori Hokokai Foundation, and Toyota Riken Scholar.
\end{acknowledgments}

\end{document}